\documentclass[12pt,preprint]{aastex}

\shorttitle{Near-Infrared Observations of RR Lyrae variables in M92}
\shortauthors{Del Principe et al.}

\begin{document}


\title{Near-Infrared Observations of RR Lyrae variables in Galactic 
Globular Clusters: I. The case of M92}


\author{M. Del Principe\altaffilmark{1} and A.M. Piersimoni\altaffilmark{1}}
\affil{INAF - Osservatorio Astronomico di Collurania, Via M. Maggini,
64100 Teramo, Italy}
\email{milena@te.astro.it, piersimoni@te.astro.it}

\author{G. Bono \altaffilmark{2} and A. Di Paola\altaffilmark{2}}
\affil{INAF - Osservatorio Astronomico di Roma, Via di Frascati, 33,
00040 Monte Porzio Catone, Italy; }
\email{bono@mporzio.astro.it, dipaola@mporzio.astro.it}

\author{M. Dolci\altaffilmark{1}}
\email{dolci@te.astro.it}

\and

\author{M. Marconi\altaffilmark{3}}
\affil{INAF - Osservatorio Astronomico di Napoli, via Moiariello 16, 
80131 Napoli, Italy}
\email{marcella@na.astro.it}




\begin{abstract}
We present near-infrared J,H, and K-band time series observations of the Galactic Globular 
Cluster (GGC) M92. On the basis of these data, we derived well-sampled light curves for 
eleven out of the seventeen cluster RR Lyrae variables, and in turn, accurate mean 
near-infrared (NIR) magnitudes. The comparison between predicted and 
empirical slopes of NIR Period-Luminosity (PL) relations indicates a very good agreement.
Cluster distance determinations based on independent theoretical NIR $PL$ relations 
present uncertainties smaller than 5\% and agree quite well with recent distance estimates 
based on different distance indicators. We also obtained accurate and deep 
NIR color-magnitude diagrams (CMDs) ranging from the tip of the Red Giant Branch (RGB) down 
to the Main Sequence Turn-Off. We detected the RGB bump and the NIR luminosities of this 
evolutionary feature are, within theoretical and empirical uncertainties, 
in good agreement with each other.
\end{abstract}





\section{Introduction}

RR Lyrae stars are the cornerstone of Population II distance indicators and
countless theoretical and observational investigations have been devoted to
the intrinsic accuracy of RR Lyrae distance determinations (Caputo et al. 2000;
Layden 2002; Cacciari \& Clementini 2003; Walker 2003).
However, we are still facing the problem that distance measurements based on
different methods present discrepancies larger than the estimated total
error budget (see e.g., Carretta et al. 2000; Bono 2003). This might be 
suggestive of still poorly constrained systematic errors. Current predictions
concerning the luminosity of Horizontal Branch (HB) models inside the RR Lyrae
instability strip bracket approximately 0.1 dex in luminosity. The difference
is due to the adopted input physics which is not well-established yet (Cassisi et al.
1998; Pietrinferni et al. 2004, and references therein).

Notwithstanding this, one of the most widespread methods to estimate RR Lyrae distances
to GGCs and Local Group galaxies is the correlation between the visual magnitude and the 
metal abundance ($M_V \propto [Fe/H]$). This method is in principle very robust, but 
presents several potential drawbacks:  {\em i)} a (strong) dependence on evolutionary 
effects (Demarque et al.  2000; Castellani 2003);
{\em ii)} a possible nonlinear dependence on the metallicity (Castellani, Chieffi, \& 
Pulone 1991; Caputo et al. 2000); {\em iii)} a non-trivial dependence on the reddening 
correction and on uncertainties affecting abundance measurements (Dall'Ora et al. 2004).
On the other hand, empirical evidence dating back to the 1990s indicates that 
$K-$band NIR observations of RR Lyrae stars appear marginally affected by deceptive
errors. In particular, $K-$band data of RR Lyrae stars, when compared with
optical $B,V-$band data, present: {\em  i)} a smaller dependence on reddening
uncertainties; {\em ii)} a smaller dependence on metal abundance; {\em iii)}
a luminosity amplitude which is approximately a factor of 4 smaller than in the
$B-$band. This means that with a limited number of phase points, the time-averaged
$K$ magnitude can be estimated with very good accuracy, since the light curve in this
band is almost sinusoidal. Moreover, and even more importantly,
Longmore et al. (1986, 1990) demonstrated, on the basis of K-band photometry for 
a good sample of GGCs, that cluster RR Lyrae do obey to a well-defined $K$-band
PL relation ($PL_K$). This finding has been soundly confirmed by Butler (2003) 
and by Storm (2004) for RR Lyrae in M3 and in IC4499. The physical basis of the 
$PL_K$ relation has been widely discussed by Bono et al. (1999; 2003). More 
recently, theoretical investigations (Bono 2003; Catelan, Pritzl, \& Smith 2004) 
have also emphasized the existence of well-defined PL relations in $J,H$ bands. 
However, observations lag theoretical predictions, and indeed we still lack 
accurate mean $K$-band magnitudes of cluster RR Lyrae, since the current 
sampling of the light curves is rather poor, typically of the order of 
4-6 phase points (Longmore et al. 1990). Moreover, $J,H$ measurements are only 
available for a limited sample of field RR Lyrae stars (Fernley et al. 1987; 
Carney et al. 1995). In order to fill this observational gap 
and to validate current predictions concerning the $J,H,K$-band $PL$ relations, 
we have undertaken a long-term project aimed at improving the current database 
of NIR observations of cluster RR Lyrae stars (Dall'Ora et al. 2004). We selected 
a dozen of northern and southern GGCs which cover a wide metallicity range and host 
a sizable sample of RR Lyrae stars.

In this investigation, we present the results for the GC M92, one of the most metal-poor 
and oldest clusters in our sample. This cluster has often been a fundamental target to
investigate the absolute age of GGCs and the chronology of the Galaxy
(Grundahl et al. 2000; VandenBerg 2000; Gratton et al. 1997; Carretta et al. 2000,
and references therein). Therefore, accurate and independent distance determinations
of this cluster can play a crucial role in the improvement of age estimates
(Renzini 1992). In the next section, we discuss the empirical data as well as the
reduction strategy adopted to perform the NIR photometry. In \S, 3 we present
optical and NIR CMDs, together with a detailed discussion concerning the key evolutionary 
features. In \S 4, we discuss NIR light curves for the sample of RR Lyrae stars in M92 
and compare empirical and theoretical NIR PL relations, while \S 5 deals with the new 
cluster distance determinations based on the NIR $\mathrm{PL}$ relations and the 
comparison with similar estimates based on various distance indicators. A brief 
summary and the conclusions are outlined in the final section.

\section{The observational scenario}


\subsection{The cluster}



The GGC M92 (NGC 6341) is very metal-poor ([Fe/H ]=$-2.24$, Zinn 1985) and is located
at a distance from the Galactic center of $R_{GC}$=9.1 kpc, and $Z_{GC}$=4.3 kpc above
the Galactic plane (Harris 1996). Therefore, it is a very good target to investigate its 
stellar content. The most recent wide-field CCD photometry was obtained in V and I bands 
with the CFH12K mosaic camera available at the $3.6\,m$ Canada-France-Hawaii Telescope
(CFHT) by Lee et al. (2003). The data presented in this paper cover a sky area significantly
larger than the tidal radius of M92. The CMD extends from the bright region down to
$5\,$ mag fainter than the Main Sequence Turn-Off. NIR photometry ($J,H,K$) of M92 
was collected
by Davidge \& Courteau (1999), who used the CFHT Adaptive Optics Bonnette for obtaining
high angular resolution JHK images of the cluster center. NIR ($J,K$) observations of giant 
stars have been recently provided by Valenti et al. (2004; hereinafter VFPO) using the NIR
camera available at the Telescopio Nazionale Galileo (TNG) to investigate evolutionary
properties of cluster RGB stars. Finally, optical (WFPC2, Piotto, Cool, \& King 1997)
and NIR (NICMOS, Lee et al. 2001) photometry have also been collected with 
the Hubble Space Telescope.

M92 hosts a sample of 21 variable stars (Sawyer-Hogg 1973; Clement et al. 2001), 
17 of them are RR Lyrae stars, 2 are SX Phoenicis,
and one is a BL Herculis, i.e. a low-mass evolved HB star (Bono et al. 1997).
A sizable sample of 60 bright RG stars, including also three candidate variables
detected by Walker (1955), were checked for variability by Welty (1985), but 
none was confirmed variable. During the last few years, the sample of 
RR Lyrae stars has
been investigated more in the optical than in the NIR bands (Carney et al. 1992;
Cohen 1992;  Cohen \& Matthews 1992; Storm, Carney \& Latham 1992, hereinafter SCL; 
Storm et al. 1992; Kopacki 2001; Piersimoni et al. 2003; Tuairisg et al. 2003). 
In particular, NIR observations for 3 RR Lyrae stars have been collected by 
Cohen (1992) and by Storm, Carney, \& Latham (1994) to estimate the distance 
to M92 using the Baade-Wesselink method.

In the following, we adopt the metal-abundance ($[Fe/H]=-2.25$) and the
$\alpha$-element enhancement ($[\alpha/Fe]=0.30$) given by Salaris \& Cassisi
(1996, see also Sneden, Pilachowski, \& Kraft 2000). By adopting these values
and the global metallicity ($[M/H]$) relation provided by Salaris, Chieffi, \&
Straniero (1993), we obtain for M92 a global metallicity $[M/H]=-2.04$.

\subsection{Observations and data reduction}

The NIR data for M92 have been collected using the AZT24 $1.1\,m$ telescope of
the Campo Imperatore (L'Aquila, Italy) observatory, located at an altitude of $2150\,m$ near 
the Gran Sasso D'Italia mountain (Di Paola 2003).
The AZT-24 is equipped with the NIR camera SWIRCAM (Brocato \& Dolci 2003), 
which is based on  a 256 x 256 HgCdTe "PICNIC" array, sensitive in the wavelength range 
$1-2.5\,\mu m$, provided by Rockwell Science Corp. The camera is provided with
the standard wide-band filters $J,H,K,K'$ plus additional narrow-band filters and grisms.
The focal plane scale is $1.04 \,arcsec/pix$, for a total field-of-view larger than 
$4.4 \times 4.4 \,arcmin^2$. The FWHM peaks around $2.1\, arcsec$ and
typically ranges from 1.8 to $2.8\, arcsec$, with the best data at $1.6\, arcsec$.
The photometric measurements of M92 were performed in the standard JHK bandpasses. 
We observed two different fields partially overlapped by 1x1 $arcmin^2$ to adopt the same
absolute calibration. The fields were chosen to observe the largest number of
RR Lyrae stars (11/17). The observations for M92 were carried out during
12 nights between June and September 2002. The typical exposure times range
from $2.5$ to $7.5\, min$ for the $J$-band, from $2.5$ to $10\, min$ for
the $H$-band, and from $2.5$ to $12.5\, min$ for the $K$-band. The individual
exposures were split in a number of sub-exposures ranging from 5 to 15.
For each sub-exposure, we dithered the images by 10 pixels both in X and in Y
direction. Target and sky frames were alternatively collected, at a distance 
of 9 arcmin from each other.

All the images were pre-processed using a pipeline called PREPROCESS, developed
by one of us (A. Di Paola). Median sky frames were subtracted from each target 
image before flat-fielding and co-adding the dithered images. Photometry was
performed using DAOPHOT/ALLSTAR (Stetson 1987). In order to improve the 
photometric accuracy along
the light curves we performed several tests using different techniques to stack
the images, different criteria to select PSF stars across the individual frames,
as well as different reduction strategies to perform the photometry over the
entire data set. We found that the best intrinsic accuracy can be achieved
by stacking at least 4 consecutive images, due to the low S/N ratio of
single images. We selected a reasonable number ($\sim 40$) of uniformly spaced
PSF stars, together with a constant PSF across each frame. To improve the accuracy
of individual measurements and the limiting magnitudes, we used a few optical 
$I$-band images to derive a more complete star catalog ($\sim$ 20000 stars).
These images were kindly provided by F. Grundahl (2004, private communication),
and have been collected at the Nordic Optical Telescope in La Palma. Finally, we
reduced NIR ($J$, $H$, $K$) and optical $I$-band frames simultaneously with 
DAOPHOT/ALLFRAME (Stetson 1994; Dall'Ora et al. 2004).

The standard field of P138-C measured by Persson et al. (1998) was observed  
10 times at about the same air mass ($\Delta X < 0.02$) of the observations 
on M92. In the calibrating equations, we neglected the extinction
with the air mass, because the extinction coefficients measured at Campo Imperatore
are $<0.1$ for all the $JHK$-bands. We compared the calibrating equations derived
for several nights with standard star observations and the accuracy of the 
NIR calibrations is of the order of 0.03 mag. Current calibrations were also 
compared with the pioneering work by Cohen, Frogel, \& Persson (1978) for M92. 
Four stars are in common with their photometric catalogue and the average 
difference in magnitude is: $\Delta J=0.015 \pm 0.014$ mag, 
$\Delta H=0.004\pm0.030$ mag, and $\Delta K = 0.012 \pm 0.013$ mag 
(our magnitudes are brighter). Therefore, the two data sets are, within 
the errors, fully consistent with no systematic offset.

The comparison between current NIR photometry and $J,K$ data collected by VFPO 
shows that the two data sets agree quite well, and indeed on a star-by-star basis 
the difference in magnitude is: $\Delta J=-0.09 \pm 0.05$ and $\Delta K=-0.07\pm 0.05$ mag
(with our magnitudes brighter). The large spread is due to the fact that our images and 
VFPO images overlap in a small $(1 \times 2\,arcmin)$ region located close to the cluster 
center, and we have roughly a dozen of common stars.

\section{The Color-Magnitude diagrams}

The NIR CMDs and the NIR Luminosity Functions (LFs) are powerful diagnostics to
investigate RG stars in GCs, since they allow us to constrain the accuracy and 
reliability
of evolutionary predictions concerning low-mass stellar structures (Ferraro et al. 2000,
and references therein). Note also that RG stars are the brightest cluster objects in NIR
bands, and therefore the unresolved background population marginally affects their photometry.
This means that high signal-to-noise NIR measurements of RG stars can be obtained close
to the cluster center, i.e. the regions more affected by severe crowding in optical
bands. Moreover, since the NIR survey of $\sim 30$ GGCs by Frogel, Cohen, \& 
Persson (1983), 
based on a single-channel detector and aperture photometry, it became clear that $V-K$ 
colors of RG stars are excellent effective temperature indicators. In spite of these 
indisputable advantages, the number of GGCs for which accurate, and deep NIR CMDs 
are available, is rather limited.

\begin{figure}
\epsscale{1}
\plotone{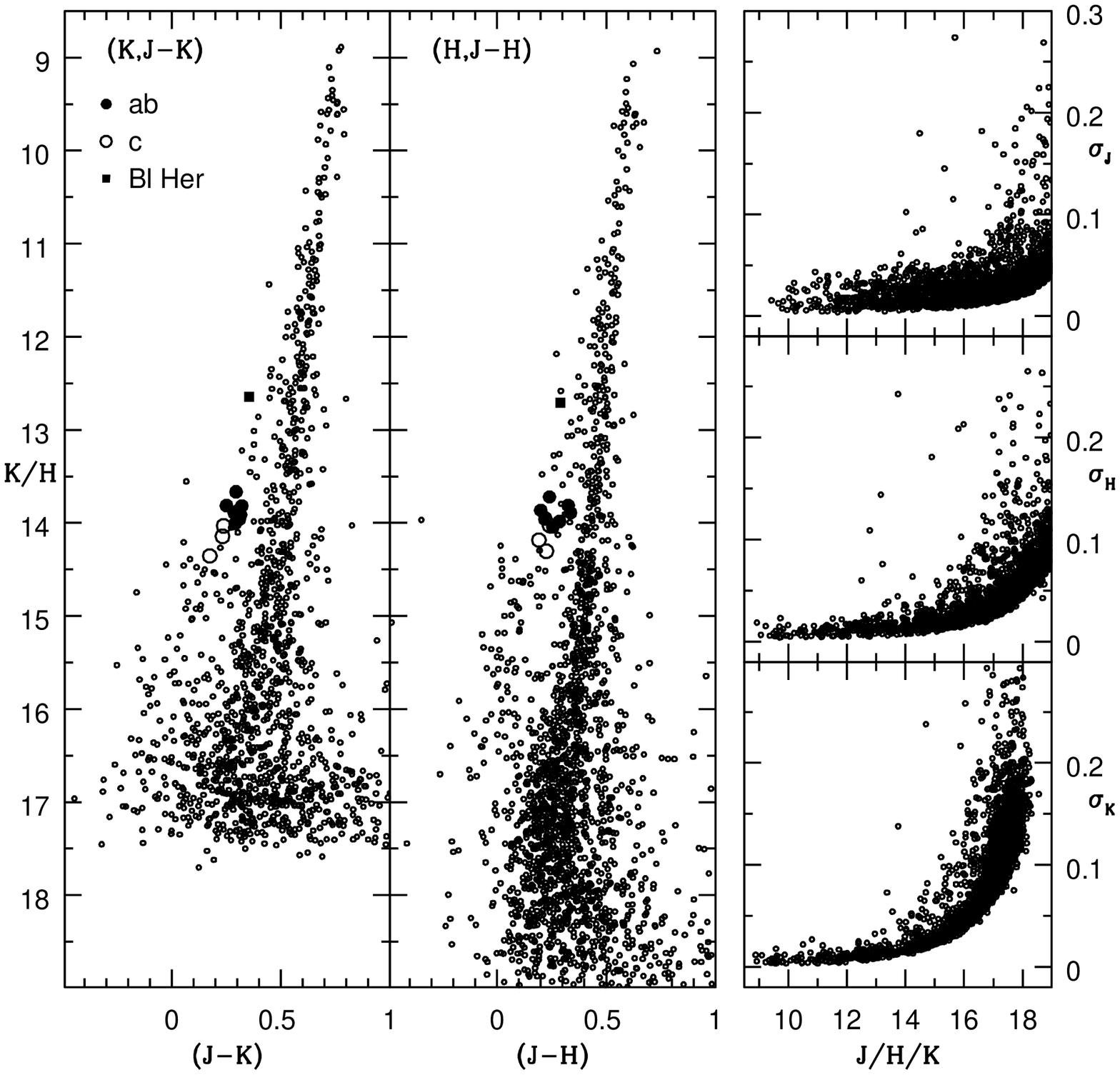}
\caption{NIR Color-Magnitude Diagrams of M92 in $(K, J-K)$ (left panel), and 
in $(H, J-H)$ (middle panel). Different symbols mark the location of variable 
stars in our sample. 
The right panels show from top to bottom the intrinsic photometric errors in 
$J,H,K$ measurements as a function of the magnitude. See text for more 
details.\label{cmdjhk}}
\end{figure}

Figure~\ref{cmdjhk} shows the $K$,($J-K$) and the $H$,($J-H$) (left and middle panel) 
CMDs of M92, together with the intrinsic photometric errors vs magnitudes (right panels).
A glance at the CMDs discloses the sizable sample ($\approx 1000$) of cluster stars
which have been measured. It is noteworthy that in the $H$,($J-H$) CMD current photometry
approaches the Turn-Off (TO) region and that the intrinsic photometric error is of the
order of 0.02-0.03 mag at the magnitude typical of RR Lyrae stars (open and filled 
circles).

\begin{figure}
\plotone{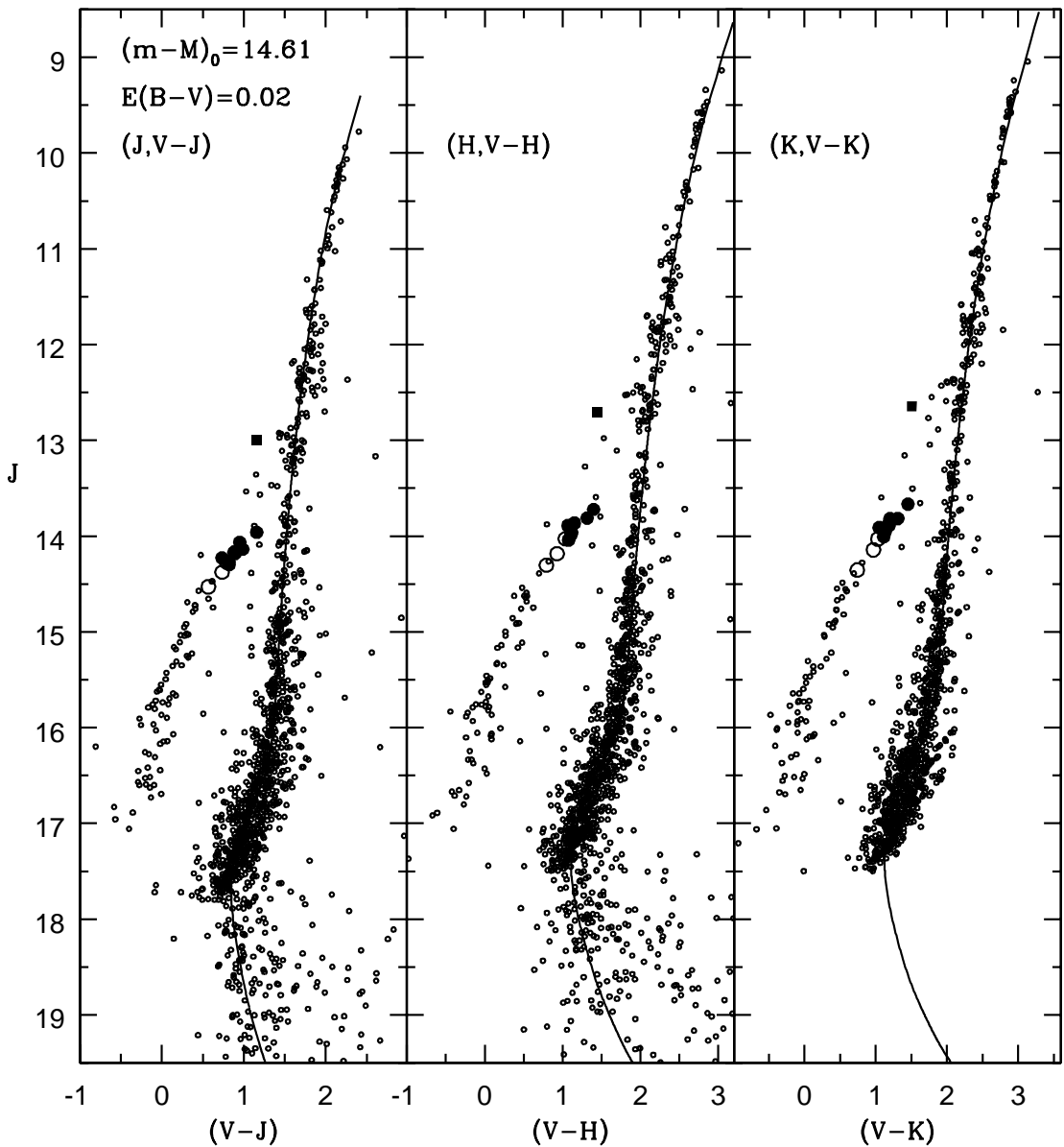}
\caption{Optical and NIR CMDs of M92 in $(J, V-J)$ (left panel), in 
$(H, V-H)$ (middle panel), and in $(K, V-K)$ (right panel). The solid 
line shows a 12 Gyr isochrone.  The adopted true distance modulus and 
cluster reddening are $(m-M)_0 =14.61$ mag and $E(B-V)=0.02$ mag as 
labeled in the figure.\label{optir}}
\end{figure}

In order to combine visual and NIR magnitudes, we adopted the V-band photometry for M92
provided by Kopacki (2001), because this is the most recent and complete optical
investigation of variable stars in M92. This choice offers the advantage of having 
homogeneous $(V-K)$ colors for both variable and static stars. Figure~\ref{optir}
shows the $J,(V-J)$, $ H,(V-H)$ and $ K,(V-K)$ CMDs based on the V photometry, kindly
provided by Kopacki, and current NIR data. Note that the faint limit of the Kopacki's
optical photometry is $\sim$ 18 mag, which is brighter than the turn-off point. However, 
current NIR photometry in $J$ and $H$ bands attains fainter limiting magnitudes. 
To overcome this problem, we also cross-correlated the $J,H$ catalogues with the deeper 
V-band photometry for M92 provided by Rosenberg et al. (1999). Detailed checks on the 
stars in common in the two V-band catalogues ($\approx 500$ stars) show that they 
agree very well, and indeed the mean difference is of the order of $0.02\pm 0.04$ mag.
The main evolutionary features of the optical/NIR CMDs are the following:

\begin{itemize}
\item{the RGB is well-populated and roughly covers the entire RGB extension.
Empirical data range from the RGB tip down to the sub-giant branch $K\approx 17$ mag.}

\item{The sample of RGB stars appears large enough - when compared with the old analysis
of Frogel, Cohen, \& Persson (1983) - to provide a reliable estimate of the RGB tip.} 

\item{Data plotted in Fig.~\ref{optir} disclose a well-defined Horizontal Branch (HB)
in the three CMDs. Moreover, hot HB stars brighter than $J\approx H\approx K\approx 16$ mag
display a well-defined slope when moving from hotter to cooler effective temperatures. This
trend is expected and is caused by the strong increase in the $K$-band bolometric correction
towards cooler effective temperatures (see Fig. 1 in Bono 2003). Interestingly enough,
this empirical evidence brings forward that $K$-band magnitudes of hot HB stars can also
be adopted to estimate cluster distances (Cassisi et al. 2005, in preparation).}
\end{itemize}

Fig.~\ref{optir} also shows a 12 Gyr isochrone for Z=0.0001, and Y=0.245\footnote{The
isochrone was retrieved from the WEB site http://www.te.astro.it/BASTI/index.php
and is based on updated evolutionary models recently provided by Pietrinferni et
al. (2004).}. The isochrone was plotted by using a cluster reddening of E(B-V)=0.02 mag
and the true distance modulus obtained by using the $PL_{JHK}$ relations (see \S 4.2).
Extinction parameters for both optical and NIR bands have been estimated using the
standard extinction model of Cardelli et al. (1989). Data plotted in this figure
disclose that a cluster age of $\sim 12$ Gyr nicely accounts for stars located
around the TO region. This age estimate is, within the errors, in good agreement
with age determinations provided by Salaris \& Weiss (2002), Grundahl et al. (2000), 
and by Carretta et al. (2000)  on the basis of deeper and more accurate
visual CMDs.

\subsection{The RGB bump and tip} \label{sect1}
\begin{figure}
\epsscale{0.7}
\plotone{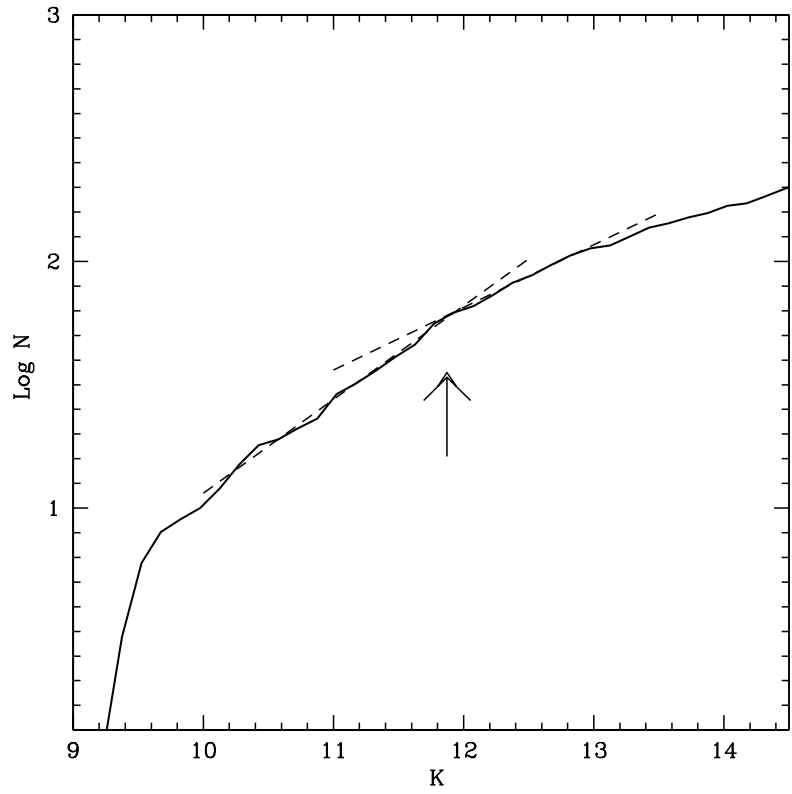}
\caption{The cumulative $K$-band RGB LF. The arrow marks the location of 
the RGB bump.\label{KLF}}
\end{figure}

The RGB Luminosity Function (LF) is a very robust observable in order to check the
accuracy of the inner structure of RGB evolutionary models. In particular, the RGB LF
can supply tight constraints upon the chemical stratification inside these stars
(Renzini \& Fusi Pecci 1988). The hydrogen stratification encountered by the thin
H-burning shell affects the evolutionary rate, and in turn, the star counts along
the RGB. In particular, the RGB bump marks the evolutionary phase during which the 
H-burning shell crosses the chemical discontinuity left over by the convective envelope 
soon after the first dredge-up. Dating back to its first detection in 47 Tuc (Lee 1977), 
several theoretical and observational investigations have been focussed on the RGB bump  
(Ferraro et al. 1999; Zoccali et al. 1999, and references therein). Owing to the 
anti-correlation of the RGB bump luminosity with the stellar metallicity, the detection 
of this feature is much easier in metal-intermediate and in metal-rich GGCs. This is 
the reason why the RGB bump has only been detected in a few metal-poor GCs. Using 
both the differential and the cumulative K-band LF (see Fig.~\ref{KLF}), we detected 
the RGB bump at $K_{bump}=11.90\pm0.08$ mag.

\begin{figure}
\epsscale{0.8}
\plotone{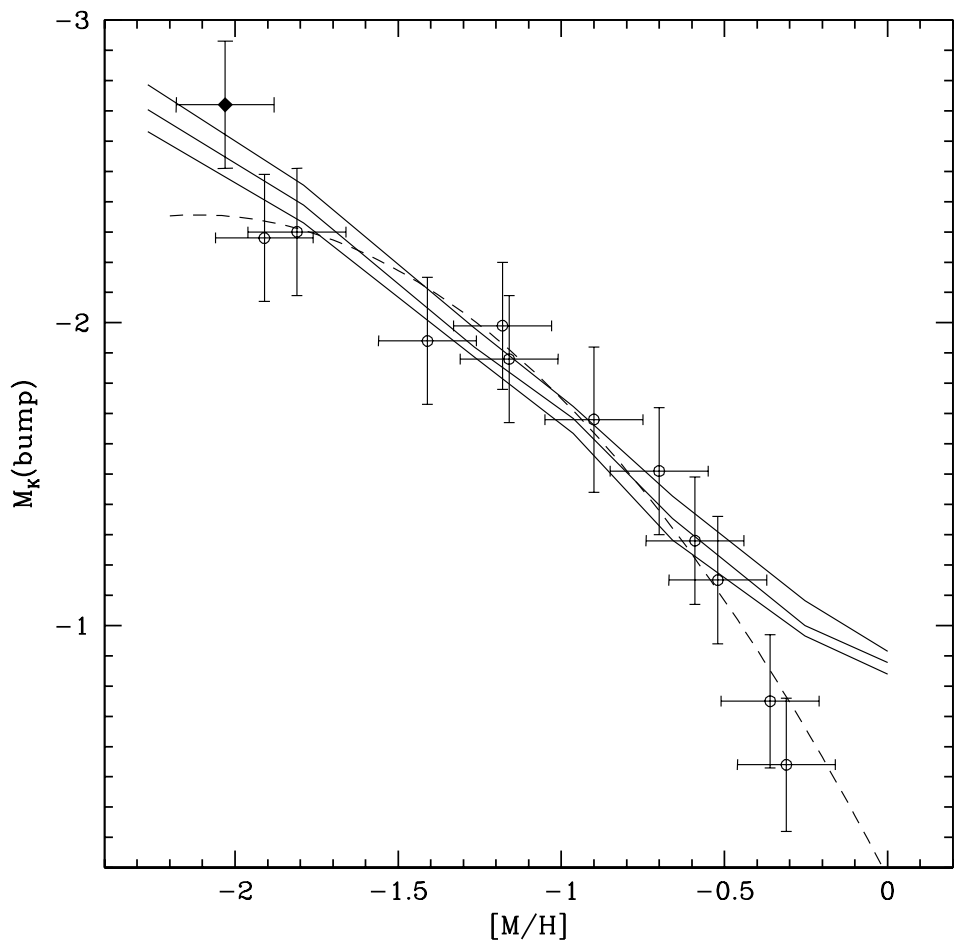}
\caption{Comparison between predicted absolute $K$ magnitude of the RGB bump as a function of 
the global metallicity (Pietrinferni et al. 2004) and observational measurements 
(Ferraro et al. 2000; VFPO). The diamond marks the location of the RGB bump in M92. 
The different solid lines display predictions for different assumptions on the cluster 
age: 12 Gyr (top line), 14 Gyrs (middle line), and 16 Gyr (bottom line). The dashed line 
is the fit to the empirical measurements provided by Valenti et al. (2004b).\label{bumpteo}}
\end{figure}

The RGB bump in M92 has also been provided by VFPO, who found a value
$K_{bump}=12.40\pm0.05$ mag. This estimate is $0.5$ mag fainter than 
current determination.
The difference is larger than the estimated intrinsic photometric errors and the
measured systematic offset between the two data sets. The difference might be 
due to the different size of RGB star samples. A glance at the data plotted 
in fig. 1 of VFPO suggests that their 
RGB is not very well populated, and thus their estimate of the error affecting the 
RGB Bump luminosity could have been underestimated. To further constrain the plausibility 
of the current $K$-band RGB bump detection, we performed a detailed comparison with updated 
evolutionary predictions (see Fig.~\ref{bumpteo}) provided by Pietrinferni et al. (2004).
In order to estimate the absolute K magnitude of the RGB bump in M92, we adopted
the true distance modulus based on the RR Lyrae $PL_{J,H,K}$ relations (see \S 4.2).
In order to perform an exhaustive comparison between theory and observations, we also
included the RGB bump estimates provided by Ferraro et. (2000) and by VFPO.
Data plotted in Fig.~\ref{bumpteo} indicate that a fairly good agreement does
exist between updated theoretical models (solid lines) and observations (see also
Cassisi \& Salaris 1997, VFPO, and references therein). However, current $M_{K}(bump)$
measurement of M92 appears at variance with the empirical trend (dashed line) found by
Valenti et al. (2004a,b). Finally, we note that the observed $K_{bump}$ is located at
$(V-K)_0=2.27$. This means $V_{bump}$=14.17 mag and, by adopting the same true distance modulus,
$M_{V}(bump)\approx -0.5$ mag, a value which agrees with theoretical predictions. Stellar isochrones 
provided by Pietrinferni et al. (2004) predict, for an age of 12 Gyr, $M_{V}(bump)\approx -0.42$ mag.

The current sample of RGB stars appears, in the brighter portion of the RGB, large enough 
to estimate the brightness of the RGB tip. If we assume that the brightest and coolest 
RGB star is at the RGB tip, one derives that $K_{tip}=8.95\pm0.2$ mag. By using the NIR 
calibration of the RGB tip provided by Bellazzini et al. (2004), i.e. 
$M_K(tip)=-(0.64\pm0.12)[M/H]-(6.93\pm0.14)$ and by adopting for M92 a global 
metallicity $[M/H]=-2.04$, we obtain $M_K(tip)=-5.62\pm 0.14$ mag. 
Note that if we adopt the true distance modulus based on the RR Lyrae 
$PL_{J,H,K}$ relations, we find that the absolute $K$ magnitude of the RGB tip is 
$M_K=-5.67\pm0.2$ mag. Therefore, the two independent estimates are, 
within the errors, identical.

\section{The RR Lyrae variables}

\begin{figure}
\epsscale{1}
\plotone{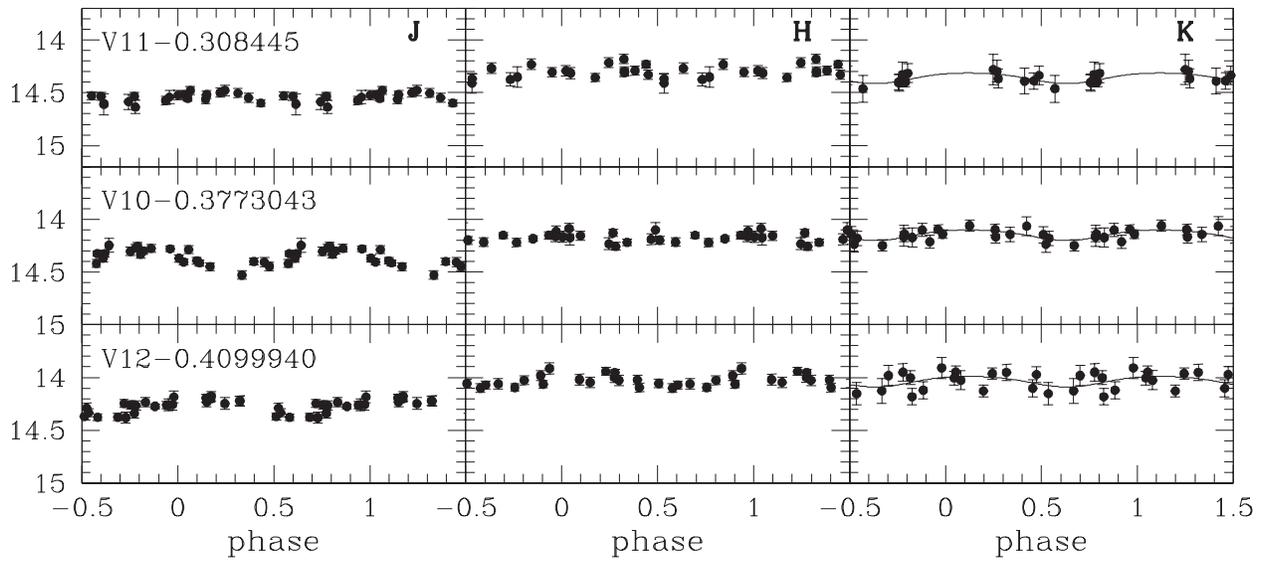}
\caption{$J$ (left), $H$ (center) $K$-band (right) light curves for first overtone 
RR Lyrae stars in M92. The labels in the left panels show the variable identification 
and the pulsation period (days). The solid lines in the right panels display the 
adopted K-band template curve. See text for more details.\label{rrc_curve}}
\end{figure}

\begin{figure}
\epsscale{1}
\plotone{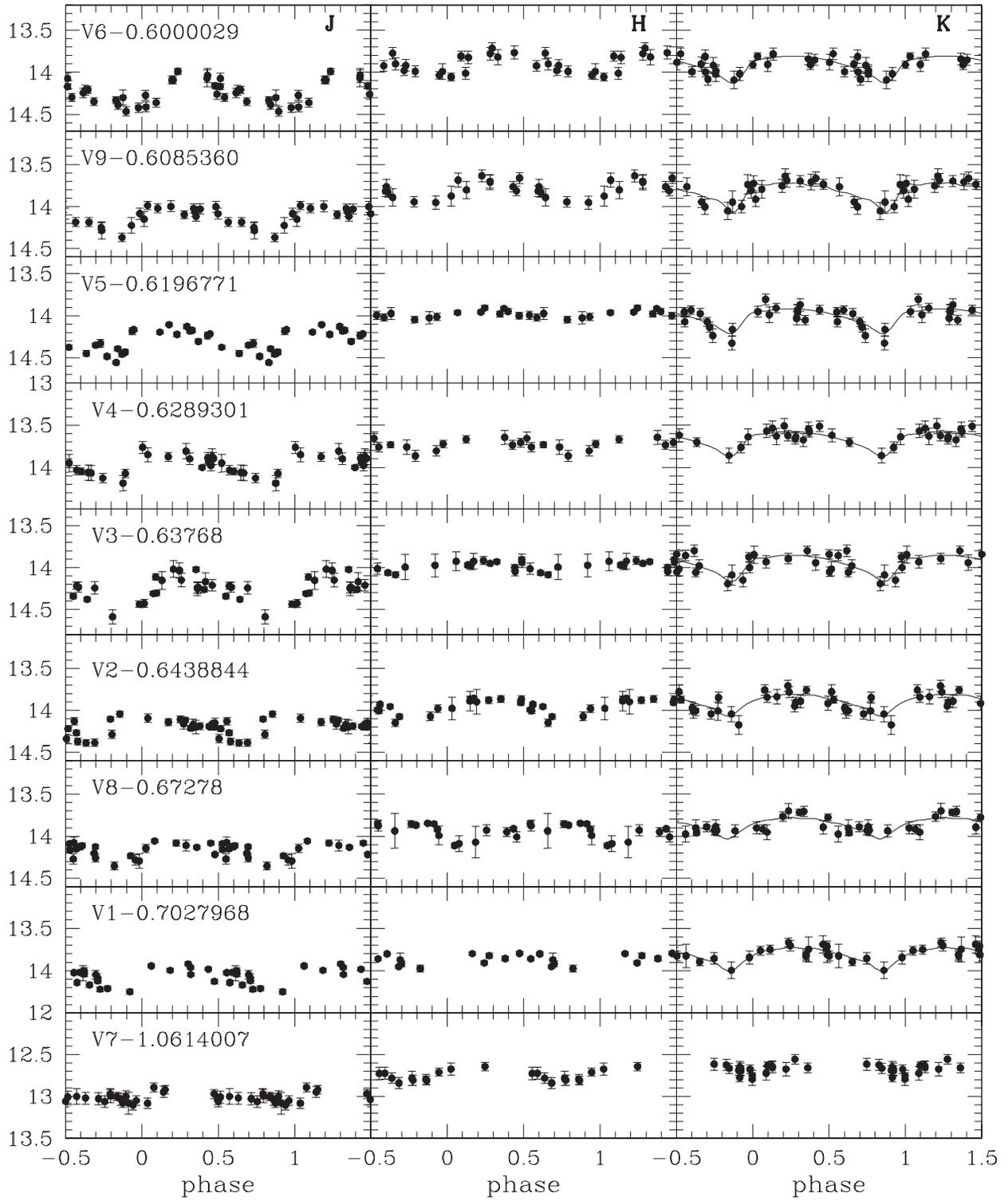}
\caption{Same as Fig. \ref{rrc_curve}, but for fundamental RR Lyrae stars and 
the BL Her variable (V7, bottom panels).
\label{rrab_curve1}}
\end{figure}

Current NIR observations allowed us to derive accurate light curves for 11 out the 17
RR Lyrae stars and for the BL Her. The light curves in the $J,H,K$ bands for first
overtone variables are shown in Figure~\ref{rrc_curve}, while the light curves for
the sample of fundamental RR Lyrae stars and the BL Her (V7) are displayed 
in Figure~\ref{rrab_curve1}.
Individual $J,H,K$ measurements are listed in Table~\ref{tabrrdata}: for each 
star we give the identification number, the Heliocentric Julian Day (HJD) of 
the observations, and the corresponding $J,H,K$ magnitudes together with 
intrinsic photometric errors. We have determined the mean $J$ and $H$ magnitudes 
of RR Lyrae by fitting the individual phase points measured by ALLFRAME with a 
cubic spline. The typical errors on the mean magnitudes are of the order 
of $0.02$ mag. On the other hand, the fit of the individual $K-$band phase 
points was performed using a template curve. This approach further improves
the intrinsic accuracy of mean $K$ magnitudes, because: {\em i)} it avoids 
the spurious fluctuations introduced by the binning of individual phase 
points; {\em ii)} it allows
a better propagation of individual photometric errors upon the mean magnitude.
This method was originally developed by Jones, Carney \& Fulbright (1996), and 
provides the mean $K$-band magnitude of RR Lyrae stars by fitting an empirical
template light curve to a few phase points. They provided five different templates 
according to the pulsation mode (fundamental, $RR_{ab}$; first overtone $RR_{c}$) 
and to the $B$-amplitude of the variable. Following this approach and by using the 
$V$-amplitudes provided by Kopacki (2001), we computed the mean $K$ magnitude for 
each star by using the template curve. Note that current mean $J,H,K$ 
magnitudes have been estimated as intensity averages and then transformed into 
magnitudes (Bono, Caputo, \& Stellingwerf 1995). 
Moreover, the $V$-amplitudes were scaled 
into $B$-amplitudes using the empirical relations provided by Jones et al. (1996).

We already mentioned that NIR $J,K$ photometry for the RR Lyrae stars in M92 was
only collected by Storm, Carney, \& Latham (1992, hereinafter SCL) and by Cohen \& 
Matthews (1992, hereinafter CM).
For the sake of comparison, Table~\ref{tabrrconfronti} gives mean $K$ and $J$ 
magnitudes for the variables in common with the quoted authors. The agreement, 
for variables V1 and V3, among independent measurements is better than 0.02 mag. 
For variable V6, the agreement with CM is better than 0.01 in the $K$-band, but 
we find a discrepancy of $\sim 0.08$ mag in the $J$-band. This discrepancy has 
no immediate explanation. 

The pulsation parameters for RR Lyrae stars in our sample together with their 
intensity averaged mean $J,H,K,V$ magnitudes and luminosity amplitudes are 
listed in Table~\ref{tabrr}. The periods, the mean $<V>$ magnitudes, and the 
$V$-band amplitudes are from Kopacki (2001). A few variables in our sample require 
individual comments. The variable V4 has a crowded neighborhood, and indeed 
the optical 
$I$-band images show a neighbor star located at a distance $\sim1\,arcsec$ from 
the variable, i.e. a distance similar to the NIR image resolution. Therefore, its 
pulsational properties are unreliable, and in the following analysis we excluded 
this variable from our sample.
The variable V7 was classified by Kopacki (2001), at variance with the Sawyer-Hogg
catalogue (1973), as a BL Her instead of a fundamental RR Lyrae. Our NIR mean 
magnitudes confirm this cassification, since this variable is on average one 
mag brighter than typical RR Lyrae stars.

\subsection{Stellar distribution along the Horizontal Branch}

\begin{figure}
\epsscale{0.8}
\plotone{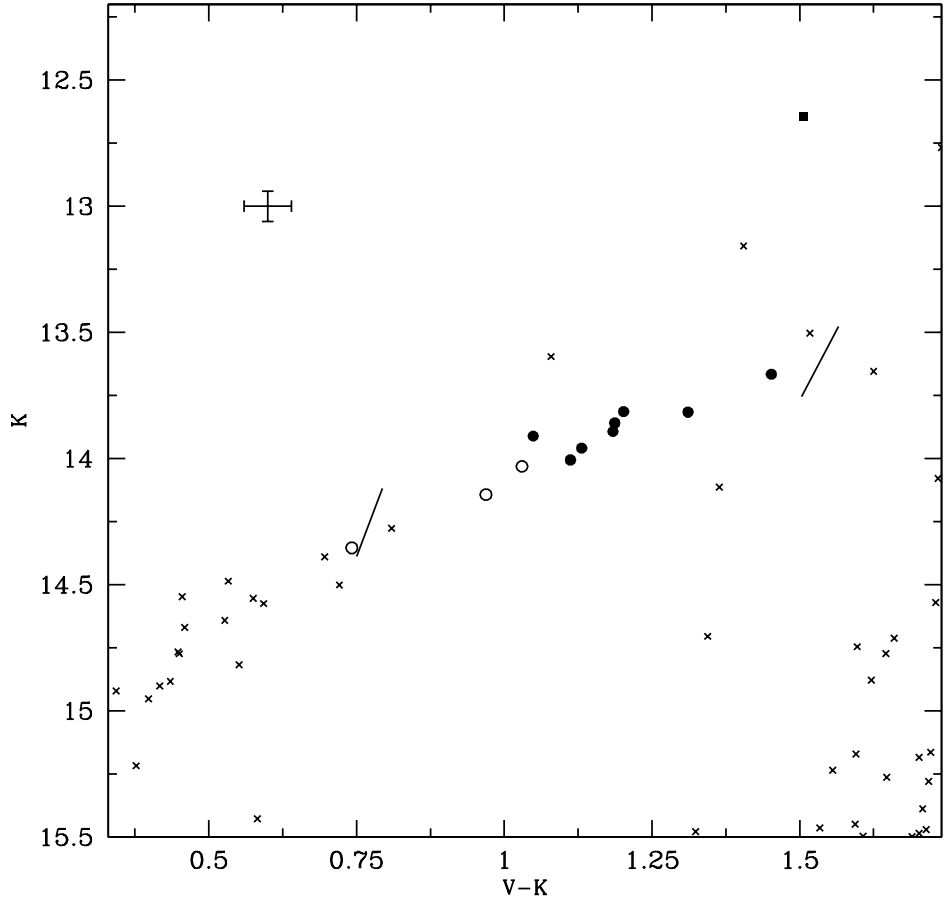}
\caption{Optical NIR CMD for both RR Lyrae stars and static HB stars. The symbols 
are the same
as in Fig.~\ref{cmdjhk}. Solid lines show the position of the theoretical instability
strip (Bono et al. 2003). The error bars in the top left corner only accounts
for observational intrinsic errors.\label{strip_m92}}
\end{figure}

On the basis of an accurate optical photometry, Carney et al. (1992) noted 
that a static star, in particular a standard star (listed by Sandage \& Walker 1966), 
was located inside the instability strip of M92.
This occurrence is not unique, and indeed Silbermann \& Smith (1995) found a similar
evidence in the instability strip of M15. The explanation suggested by Carney et al. (1992)
was that $B-V$ colors are not good effective temperature indicators.
To further investigate this point, Figure~\ref{strip_m92} shows the distribution in
the $(V-K),K$ plane of both RR Lyrae and static HB stars ($\times$ symbols) in M92.
The predicted edges of the instability strip are plotted as solid lines and are based
on RR Lyrae pulsation models provided by Bono et al. (2003). These models were
constructed by adopting metal and Helium abundance of Z=0.0001, Y=0.24, respectively.
Predictions were transformed into the observational plane using the atmosphere models
provided by Castelli \& Kurucz (2004). In the comparison between theory and
observations, we adopted the same reddening and true distance modulus adopted in the
isochrone fit.  Data plotted in this figure show the expected ranking of the
$V-K$ colors with the pulsation period, and indeed first overtone and fundamental
pulsators do not overlap. Moreover, we also found three outliers in the instability
strip. A first overtone is located outside the instability strip, while two static
stars are located inside the instability strip: one is close to the first overtone blue
(hot) edge, and one is near the fundamental red (cool) edge. Interestingly enough, 
the latter static star ($V-K=1.52$, $K=13.50$) is the star identified by 
Carney et al. (1992).
However, these objects in the $(V-K), K$ plane do not pose a severe problem, since
their location can be safely explained if we account for theoretical uncertainties
affecting both pulsation and atmosphere models (Bono et al. 1997) as well as for
empirical uncertainties (see error bars). This finding strongly supports the
hypothesis suggested by Carney et al. (1992) to explain the occurrence of
static stars inside the instability strip.  
Finally, we would like to mention that the two static stars located below/above 
the instability strip might be either photometric blends or field stars.

\subsection{The NIR period-luminosity relations}

During the last few years, a substantial theoretical effort has been devoted to
the pulsational properties of RR Lyrae stars in NIR bands. These investigations
rely on predictions based either on pulsational models (Bono et al. 2001,2002,2003) 
or on synthetic horizontal branches (Catelan et al. 2004; Cassisi et al.  2004),
and provide $\mathrm{PL}$ relations in different NIR bands.
More in detail, Cassisi et al. (2004) investigated the predicted behavior of RR Lyrae
stars as a function of the HB morphology over a wide metallicity range, namely
$(0.0001 <Z<0.006)$. Interestingly enough, they found that for a HB type\footnote{
This parameter was introduced by Zinn (1980) and is the ratio $N_B/(N_B+N_{RR}+N_R)$, where 
$N_B$ and $N_R$ are the number of HB stars bluer/redder than RR Lyrae stars and $N_{RR}$ is 
the number of RR Lyrae stars. GGCs characterized by a very blue HBs have HB 
type close to one, while GGCs with very red HBs have HB type close to zero.} = 0.90,
i.e. the HB type of M92, the $\mathrm{PL_K}$ relation does not depend on the 
metallicity and is the following:

\begin{equation}
\label{eqn1}
     <M_K> = -2.30(\pm 0.01)(LogP+0.30)-0.46(\pm 0.01) 
\end{equation}

On the other hand, Catelan et al. (2004) using a similar approach and synthetic
HB models covering a narrower metallicity range  $0.0005<Z<0.006$, found very
similar NIR $\mathrm{PL}$ relations. In particular, their relation, for HB Type=0.934 
and Z=0.0005, are the following:
\begin{mathletters}
\begin{eqnarray}
\label{eqnarray1}
   <M_J> = -1.902(\pm 0.045)LogP - 0.826(\pm 0.012)\\
\label{eqnarray2}
   <M_H> = -2.311(\pm 0.013)LogP - 1.136(\pm 0.003) \\
\label{eqnarray3}   
   <M_K> = -2.343(\pm 0.012)LogP - 1.168(\pm 0.002)
\end{eqnarray}
\end{mathletters}
The two sets of $\mathrm{PL}$ relations, for P=0.5 days, Z=0.0001, and HB Type=0.95
supply $K$ magnitudes that differ by only 0.01 mag. Therefore, we decide to use the 
NIR $\mathrm{PL}$ relations provided Catelan et al. (2004) to fit our data, though 
they do not cover the global metallicity of M92.

In order to provide two independent theoretical frameworks to be compared with empirical
data we also computed, using the same evolutionary and pulsation models
adopted by Cassisi et al. (2004), new $J,H$-band $\mathrm{PL}$ relations.
We found:

\begin{equation}
\label{eqn3}
<M_J> = -1.708(\pm 0.006)(LogP+0.30)-0.240(\pm 0.004)
\end{equation}

\begin{equation}
\label{eqn4}
<M_H> = -2.26(\pm 0.01)(LogP+0.30) -0.44(\pm 0.01)
\end{equation}

\begin{figure}
\epsscale{1}
\plotone{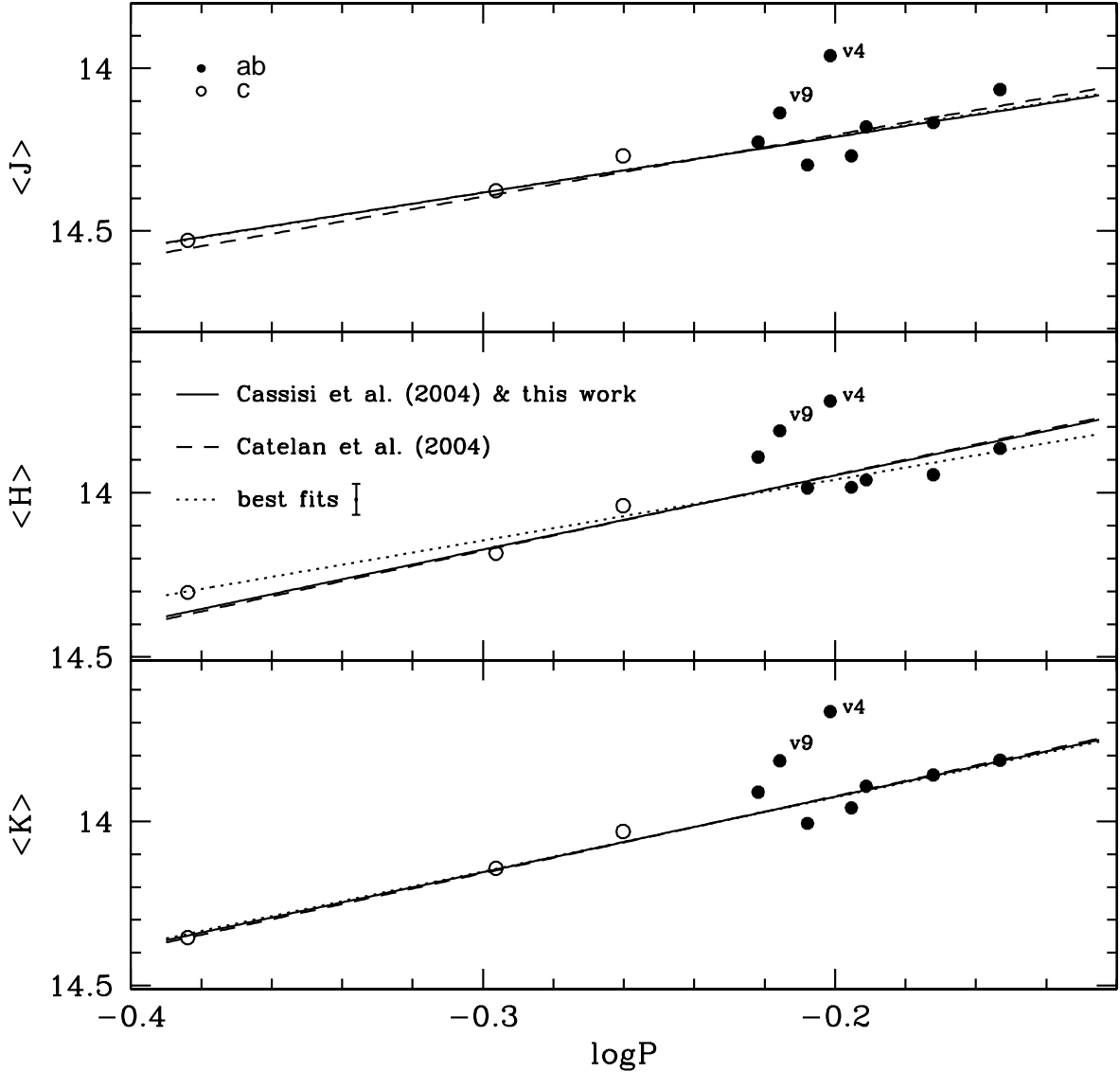}
\caption{NIR PL relations for RR Lyrae stars in M92, in $J$ (top), $H$ (middle), and 
$K$ (bottom) bands. The periods of first overtone pulsators ($RR_c$, open circles) 
have been fundamentalized, i.e. we added 0.127 to $\log P_{FO}$. Filled circles are 
fundamental RR Lyrae ($RR_{ab}$) stars. The dotted lines display fits to empirical 
data, while the solid lines the theoretical slopes provided by Cassisi et al. (2004) 
for the $K$-band, and by current investigation for the $J,H$-bands 
(see eqs.~\ref{eqn1},\ref{eqn3},\ref{eqn4}). The dashed lines show the slopes 
predicted by Catelan et al. (2004, see eqs. 2a,2b,2c). The variables V4 
and V9 were not included in the empirical fit. The error bar plotted in the 
middle panel shows the standard deviation of empirical best fits. The standard 
deviation of theoretical relations are smaller than symbol size.\label{plm92}}
\end{figure}

From an empirical point of view, there are several $PL_K$ relations available in the
literature, but a general consensus on the slope and on the zero-point has not been 
reached yet. They range from   $M_K=-1.92\times\log{P}-0.74$
(Longmore et al. 1990), to $M_K=-2.95(\pm0.10)\times\log{P}-1.07(\pm0.10)$ (Skillen et al. 1993).
In Figure~\ref{plm92}, we plot the mean $J$, $H$, and $K$ magnitudes for our sample
of RR Lyrae variables as a function of the fundamentalized period 
($\log P_F=\log P_{FO}+0.127$). The best fits to empirical data\footnote{Note that we 
did not include variable V9 in the best fit estimates, because it is located near the 
center of the cluster. Therefore, the pulsational properties of this variable are less 
accurate than for the other variables. However, if we account for this variable the best 
fits become: $<J>=-1.77(\pm0.28)LogP+13.845(\pm0.057)$, $<H>=-1.91(\pm0.37)LogP+13.558(\pm0.076)$,
 and $<K>=-2.30(\pm0.30)LogP+13.449(\pm0.061)$.} 
provide the following relations:
\begin{eqnarray}
\label{eqn5}
<J>=-1.73(\pm0.24)LogP+13.863(\pm0.049)\\
\label{eqn6}
<H>=-1.85(\pm0.25)LogP+13.590(\pm0.051)\\
\label{eqn7}
<K>=-2.26(\pm0.20)LogP+13.475(\pm0.040)
\end{eqnarray}

In each panel, the empirical best fits (dotted lines) have been overlapped to the data 
together with the predicted $K$-band $\mathrm{PL}$ relation by Cassisi et al. (2004), the 
new $J,H$-band  $\mathrm{PL}$ (see equations  \ref{eqn1},\ref{eqn3},\ref{eqn4}; 
solid lines) and the NIR $\mathrm{PL}$ relations provided by Catelan et al. (2004, 
see equations 2a,2b,2c; dashed lines). 
Data plotted in this figure disclose that predicted and observed slopes agree quite well 
in the three NIR photometric bands, thus supporting the plausibility and the accuracy of 
current pulsation and evolutionary models.

\section{Discussion}

Current findings together with similar results for cluster and field RR Lyrae stars 
(Butler 2003; Bono et al. 2003; Borissova et al. 2004; Dall'Ora et al. 2004) support 
the evidence that NIR $\mathrm{PL}$ relations can supply accurate distance estimates 
(Bono et al. 2001). Therefore, by using the new $J$-band $\mathrm{PL}$ relation 
(eq.~\ref{eqn3}) and individual mean $J$ magnitudes, we estimated the apparent 
distance modulus $(m-M)_J=<J>-M_J$ for each RR Lyrae in our sample. The weighted 
average of these estimates gives $(m-M)_J=14.62\pm0.05$ mag. 
We adopted the same approach for $H$ (eq.~\ref{eqn4}) and $K$-band measurements 
and we found $(m-M)_H=14.61\pm0.06$ and $(m-M)_K=14.62\pm0.04$ mag, respectively. 
The quoted errors account 
only for the dispersion around the mean of the individual estimates. The typical reddening value 
for M92 is $E(B-V)=0.02$ mag (Harris 1996). By using the standard extinction model by 
Cardelli et al.(1989), we found that $A_J=0.282\times A_V$, $A_H=0.190\times A_V$, and 
$A_K=0.114\times A_V$, which give absorptions smaller than 0.02 mag in NIR bands 
($A_J=0.017$, $A_H=0.012$, $A_K=0.007$).

The final error budget was evaluated by accounting for the dispersion around the mean 
of the individual estimates, for the absolute photometric calibration ($\approx0.03$),
and for the uncertainty affecting the zero-point of the predicted $\mathrm{PL}$ relation 
($\approx0.01$). By adding in quadrature these error sources, we obtain a "global" 
uncertainty of 0.05 mag for the $K$-band, 0.06, and 0.07 mag for the $J$ and $H$ band, 
respectively. By accounting for reddening corrections in the three different bands, the 
true cluster distance modulus, obtained as a weighted average of $J$, $H$, and $K$ 
estimates (see Table~\ref{tabmod}), is equal to $(m-M)_0=14.61\pm0.03$ 
mag\footnote{This distance estimate does not include the variable V9. If we account 
for this variable the true cluster distance is  $(m-M)_0=14.59\pm0.04$ mag.}.
Table~\ref{tabmod} also gives the true distance moduli, based on the NIR 
$\mathrm{PL}$ relations predicted by Catelan et al. (2004)\footnote{The variable 
V9 was not included.}. It is worth noting that the two sets of independent 
distance determinations are, within the uncertainties, identical.  

During the last few years, the discussion concerning pros and cons of different methods to
estimate the distances to GGCs has been addressed in several theoretical and empirical
investigations (Reid 1997; Pont et al. 1998; Carretta et al. 2000; Caputo et al. 2000; 
Cacciari 2003). In particular, Pont et al. (1998) and Carretta et al. (2000) using
the subdwarf Main Sequence fitting found true distance moduli for M92 which 
agree
quite well with each other, $14.61\pm0.05$ mag and $14.64\pm0.07$ mag, respectively.
To further constrain the accuracy of distance determinations to M92, we also
adopted the First Overtone Blue Edge (FOBE) method suggested by Caputo et al. (2000).
This method relies on the comparison in the Period-$V$-band magnitude plane between 
predicted and empirical $V$-band first overtone blue edge. By using this method and 
the visual magnitudes
provided by Kopacki (2001) for the three $RRc$ variables in M92, we derived a true 
distance modulus equal to $14.62\pm0.07$ mag. To provide a comprehensive scenario 
concerning distance determinations to M92, we also used the predicted HB luminosity 
provided by Pietrinferni et al. (2004). Theoretical HB models predict the luminosity 
of the Zero Age Horizontal Branch at the level of the RR Lyrae instability strip, i.e.
$\log{T_e}\sim3.85$. However, HB stars in NIR bands do show a well-defined slope.
In order to overcome this problem, we used predicted and empirical $V-K$ colors to estimate
individual effective temperatures. We selected two variables, V10 and V11, for which
we found $\log {T_e} = 3.84$ and $3.86 $ respectively. The mean K magnitude of these
two variables is: $<K>=14.24\pm0.1$ mag. The predicted absolute $K$-band magnitude of 
the ZAHB at the metallicity of M92 is equal to: $M_K=-0.393$ mag. Therefore, the true 
distance modulus is equal to $(m-M)_0=14.62\pm0.1$ mag. Note that the comparison between
the $K-$band mean magnitude of RR Lyrae stars and predicted $K-$band ZAHB brightness
at $\log {T_e} = 3.85$ appears as a very promising approach, since in this band the 
evolutionary effects also imply a change in effective temperature (Bono et al. 2001).
The seventh column of Table~\ref{tabmod} gives the distance to M92 estimated using the 
near-infrared Baade-Wesselink (BW) method for two cluster RR Lyrae stars observed 
by  Cohen (1992) and by Storm et al. (1994). The last column in Table~\ref{tabmod} 
lists the distance modulus to M92 using the empirical calibration of the luminosity 
of the RGB tip recently provided by Bellazzini et al. (2004) and shortly
discussed in Sect.~\ref{sect1}  
  
Interestingly enough, data listed in Table 5 disclose that distance determinations 
based on different methods and photometric bands agree quite well, within current 
empirical and theoretical uncertainty. This outcome applies to distance estimates 
based on both empirical and theoretical calibrations of different distance indicators. 
In particular, the agreement is better than 5\% in distance estimates based on pulsational 
and evolutionary methods (predicted NIR $\mathrm{PL}$ relations, FOBE, ZAHB) and on the 
MS fitting.  An uncertainty of the order of 10\% for the tip of the RGB is expected 
due to the limited number of stars across this evolutionary phase in metal-poor clusters. 
Distances based on the BW method present larger uncertainties. The difference might 
be due to the fact that BW distances are only based on two cluster variables.

\section{Summary and final remarks}

We collected new multiband NIR ($J,H,K$) time series data of the GGC M92.  
The main results of this investigation can be summarized as follows:

\begin{itemize}
\item{We performed an extensive and accurate NIR photometric investigation of 
RR Lyrae stars in the GGC M92. For the first time, we provided well-sampled 
$J,H,K$-band light curves for eleven cluster RR Lyrae variables.}

\item{We provided accurate estimates of NIR pulsation properties (mean magnitudes,
luminosity amplitudes). The comparison between current data and updated theoretical
prescriptions discloses a very good agreement concerning the slope of the
$J,H,K$-band $\mathrm{PL}$ relations.}

\item{By adopting mean $J$, $H$ and $K$ magnitudes for our sample of RR Lyrae stars and
by relying on theoretical $\mathrm{PL}$ relations derived in this work for the $J$ and $H$-bands 
and predicted by Cassisi et al. (2004) for the $K$ band, we estimated the distance to M92 with 
an accuracy better than 5\%. The same outcome applies if we use the NIR $\mathrm{PL}$ 
relations provided by Catelan et al. (2004).  
Our distance estimates are also in very good agreement with recent accurate and 
independent measurements based on various distance indicators. This finding
further strengthens the use of predicted NIR $\mathrm{PL}$ relations
for estimating distances to GGCs and Local Group galaxies.}
\end{itemize}

The current investigation clearly indicates the key-role of new and accurate
NIR photometry of RR Lyrae stars in GGCs in constraining their pulsational and 
evolutionary properties and to assess the reliability and accuracy of NIR 
$\mathrm{PL}$ relations as distance indicators for field and cluster 
RR Lyrae stars. Needless to say, it is quite important to extend such 
photometric analysis to other GGCs, in order to investigate the dependence 
on the metal content, on the evolutionary status, and on HB morphology 
(Cassisi et al. 2004; Catelan et al. 2004).

\acknowledgments

We are grateful to F. Grundahl for kindly providing several optical frames of M92
and G. Kopacki for sending us his VI photometry of M92. We also wish to thank 
A. Arkharov and V. Larionov, who have collected calibration frames during several 
nights and F. Caputo and S. Cassisi for many enlightening discussions. 
It is a real pleasure to acknowledge an anonymous referee for his/her 
positive comments and pertinent suggestions.
This work was partially supported by MURST (PRIN2002, PRIN2003).

\clearpage



\clearpage


\clearpage

\clearpage
\begin{table}
\begin{center}
\caption{$J,H,K$-band photometry of variable stars.\label{tabrrdata}}
\begin{tabular}{ccccccccc}
\tableline\tableline
\multicolumn{9}{c}{V1}\\
\tableline
HJD-2452000\tablenotemark{a}&  J & $\sigma_J$\tablenotemark{b} & HJD-2452000\tablenotemark{a}& H& $\sigma_H$\tablenotemark{b} & HJD-2452000\tablenotemark{a} & K& $\sigma_K$\tablenotemark{b}\\ 
\tableline
  438.5094 & 13.919 & .023 & 438.5627  & 13.857  & .029    &  431.9563  & 13.844  &  .075   \\
  438.5204 & 13.959 & .022 & 446.5131  & 13.869  & .080    &  439.0346	& 13.763  &  .054   \\
  446.4588 & 14.030 & .089 & 460.4153  & 13.794  & .019    &  446.9874	& 13.747  &  .147   \\
  460.3892 & 13.981 & .015 & 460.5087  & 13.801  & .020    &  460.9511	& 13.667  &  .051   \\
  460.4740 & 14.022 & .019 & 466.9004  & 13.917  & .036    &  465.9515	& 13.829  &  .060   \\
\tableline
\end{tabular}


\tablecomments{Table \ref{tabrrdata} is published in its entirety in the electronic 
edition of the {\it Astronomical Journal}. A portion is shown here for guidance 
regarding its form and content.}
\tablenotetext{a}{Heliocentric Julian Day.}
\tablenotetext{b}{Intrinsic photometric errors.}
\end{center}
\end{table}


\begin{table}
\begin{center}
\caption{Mean $K$ and $J$ magnitudes for the RR Lyrae variables in common with
Storm, Carney \& Latham (1992, SCL) and with Cohen \& Matthews (1992, CM).\label{tabrrconfronti}}
\begin{tabular}{cccccc}
\tableline\tableline
 ID & $<K>_{this~work}$  & $<K>_{SCL}$ & $<K>_{CM}$ &  $<J>_{this~work}$ & $<J>_{CM}$ \\
\tableline
V1  & 13.81             & 13.823      & 13.800     & 14.07              & 14.066 \\
V3  & 13.96             & 13.958      & \ldots     & \ldots             & \ldots \\
V6  & 13.91             & \ldots      & 13.897     & 14.23              & 14.149 \\
\tableline
\end{tabular}


\end{center}
\end{table}

\clearpage
\begin{table}
\scriptsize 
\begin{center}
\caption{Pulsational properties of RR Lyrae stars in our sample. The 
identification numbers are from the Sawyer-Hogg (1973) catalogue, while
periods, mean V magnitudes and V-band amplitudes are from Kopacki (2001). 
We estimated for the mean magnitudes a global error of 0.03 mag, which 
is dominated by the intrinsic error affecting the absolute zero-point 
calibration.\label{tabrr}}
\begin{tabular}{llccccccccl}
\tableline\tableline
 {ID}  & $P$ & $<J>\tablenotemark{a}$ &  $A_J\tablenotemark{b}$ &  $<H>\tablenotemark{a}$ & $A_H\tablenotemark{b}$ & $<K>\tablenotemark{a}$ & $A_K\tablenotemark{b}$ & $<V>-<K>$ & $A_V\tablenotemark{b}$ & {Type} \\
       &  (days)     & (mag)  & (mag)  & (mag) & (mag) & (mag) & (mag) & (mag) & (mag) & \\
\tableline
V1  & 0.7027968 &14.07  &0.31  &13.87  &0.18  &13.81  &0.28  &1.21  &0.86  & ab \\
V2  & 0.6438844 &14.18  &0.37  &13.96  &0.27  &13.89  &0.27  &1.19  &0.85  & ab \\
V3  & 0.63768   &14.27  &0.41  &13.98  &0.14  &13.96  &0.31  &1.13  &1.17  & ab \\
V4  & 0.6289301 &13.96  &0.38  &13.72  &0.23  &13.67  &0.29  &1.45  &0.84  & ab \\
V5  & 0.6196771 &14.30  &0.34  &14.04  &0.10  &14.01  &0.30  &1.11  &0.98  & ab \\
V6  & 0.6000029 &14.23  &0.39  &13.89  &0.26  &13.91  &0.32  &1.05  &1.09  & ab \\
V7  & 1.0614007 &13.00  &0.11  &12.71  &0.19  &12.64  &0.16  &1.51  &0.64  & Bl Her\\
V8  & 0.67278   &14.17  &0.31  &13.95  &0.22  &13.86  &0.26  &1.19  &0.67  & ab \\
V9  & 0.6085360 &14.14  &0.37  &13.81  &0.28  &13.82  &0.36  &1.31  &1.19  & ab \\
V10 & 0.3773043 &14.38  &0.23  &14.18  &0.10  &14.14  &0.11  &0.97  &0.48  & c  \\
V11 & 0.308445  &14.53  &0.18  &14.30  &0.16  &14.35  &0.11  &0.75  &0.62  & c  \\
V12 & 0.4099940 &14.27  &0.18  &14.03  &0.10  &14.03  &0.11  &1.03  &0.41  & c  \\
\tableline

\tablenotetext{a}{Intensity averaged mean $J,H,K$-magnitudes.}  
\tablenotetext{b}{Luminosity amplitudes in $J,H,K,V$-bands.}  
\end{tabular}
\end{center}
\end{table}

\clearpage
\begin{table}
\scriptsize 
\begin{center}
\caption{True distance moduli for M92 estimated by using different distance 
indicators.\label{tabmod}}
\begin{tabular}{cccccccc}
\tableline\tableline
$PL_{J}$ & $PL_{H}$ & $PL_{K}$ & $FOBE$ & MS fitting & ZAHB & BW & RGB Tip\\
\tableline
$14.61\pm0.06$\tablenotemark{a} &  $14.60\pm0.07$\tablenotemark{a}  & 
$14.61\pm0.05$\tablenotemark{a} &  $14.62\pm0.07$\tablenotemark{c}  & 
$14.61\pm0.05$\tablenotemark{d} &  $14.62\pm0.10$\tablenotemark{f}  & 
$14.47\pm0.15$\tablenotemark{g} &  $14.56\pm0.20$\tablenotemark{i} \\   
$14.63\pm0.05$\tablenotemark{b} &  $14.61\pm0.06$\tablenotemark{b}  & 
$14.62\pm0.04$\tablenotemark{b} &  \ldots                           & 
$14.64\pm0.07$\tablenotemark{e} &  \ldots                           & 
$14.60\pm 0.26$\tablenotemark{h}  & \ldots   \\
\tableline

\tablenotetext{a}{Distance determinations based on predicted RR Lyrae 
$J,H,K$-band $\mathrm{PL}$ relations (see eqs.~\ref{eqn1},\ref{eqn3},\ref{eqn4}).}
\tablenotetext{b}{Distance determinations based on predicted RR Lyrae $J,H,K$-band 
$\mathrm{PL}$ relations provided by Catelan et al. (2004, see eqs.~\ref{eqnarray1}
,~\ref{eqnarray2},~\ref{eqnarray3}).}
\tablenotetext{c}{Distance determination based on the First Overtone Blue Edge method 
(Caputo et al. 2000).}
\tablenotetext{d}{Distance determination based on the Main Sequence fitting (Pont et al. 1998).}
\tablenotetext{e}{Distance determination based on the Main Sequence fitting (Carretta et al. 2000).}
\tablenotetext{f}{Distance determination based on ZAHB models (Pietrinferni et al. 2004).}
\tablenotetext{g}{Distance determination based on the BW method (Cohen 1992).}
\tablenotetext{h}{Distance determination based on the BW method (Storm et al. 1994).}
\tablenotetext{i}{Distance determination based on the RGB Tip (Bellazzini et 
al. 2004).}
\end{tabular}
\end{center}
\end{table}

\end{document}